\documentclass{article}
\usepackage{mpgstyle}
\usepackage{amsmath,amssymb,amscd,amsthm,amsfonts,
color,
enumerate,fix-cm,layout,lipsum,
mathrsfs,
stmaryrd,
tikz,
xspace,url}
\usepackage{hyperref}
\usepackage[all]{xy}
\theoremstyle{plain}
\newtheorem{theorem}{Theorem}[section]

\theoremstyle{definition}

\theoremstyle{remark}
\newtheorem{remark}[theorem]{Remark}

\hypersetup{colorlinks=true,linkcolor=blue,citecolor=purple,filecolor=magenta,urlcolor=cyan}

\def\be{\begin{eqnarray}}
\def\ee{\end{eqnarray}}
\def\nn{\nonumber}

  \newcommand\qtTwo[0]{[2]_{qt}}
  \newcommand\qtThree[0]{[3]_{qt}}

\def\CP1{\mathbb{C}\mathrm{P}^1}

\allowdisplaybreaks[3]

\newcommand\prePositiveCrossing{
  \put(0,0){
    \put(0,0){\qbezier(0,0)(7.5,7.5)(15,15)}
    \put(15,0){\qbezier(0,0)(-4,4)(-6,6)}
    \put(0,15){\qbezier(0,0)(4,-4)(6,-6)}
  }
}

\newcommand\preNegativeCrossing{
  \put(0,0){
    \put(0,0){\qbezier(15,0)(7.5,7.5)(0,15)}
    \put(0,0){\qbezier(0,0)(4,4)(6,6)}
    \put(15,15){\qbezier(0,0)(-4,-4)(-6,-6)}
  }
}

\newcommand\arrowsUp{
  \put(15,15){\vector(1,1){0}}
  \put(0,15){\vector(-1,1){0}}
}

\newcommand\crossingPosUp{
  \prePositiveCrossing
  \arrowsUp
}

\newcommand\crossingNegUp{
  \preNegativeCrossing
  \arrowsUp
}

\newcommand\brackets[1]{{\left( #1 \right)}}
\newcommand\braces[1]{{\left\{ #1 \right\}}}
\newcommand\opor[0]{\text{ or }}
\newcommand\opand[0]{\text{ and }}

\preprintAdd{ITEP/TH-10/19}
\preprintAdd{IITP/TH-06/19}
\preprintAdd{MPTI/TH-01/19}

\title{Nimble evolution for pretzel Khovanov polynomials}

\author[b]{Aleksandra Anokhina,}
\author[a,b,d]{Alexei Morozov,}
\author[a,b,c,d]{Aleksandr Popolitov}
\bigskip

\affiliation[a]{Moscow Institute for Physics and Technology, Dolgoprudny, Russia}
\affiliation[b]{ITEP, Moscow 117218, Russia}
\affiliation[c]{Department of Physics and Astronomy, Uppsala University, Box 516, SE-75120 Uppsala, Sweden.}
\affiliation[d]{Institute for Information Transmission Problems, Moscow 127994, Russia}

\emailAdd{anokhina@itep.ru}
\emailAdd{a.popolitov@physics.uu.se}
\emailAdd{morozov@itep.ru}

\abstract{
  We conjecture explicit evolution formulas for Khovanov polynomials,
  which for any particular knot are Laurent polynomials of complex variables $q$ and $T$,
  for pretzel knots of genus $g$
  in some regions in the space of winding parameters $n_0, \dots, n_g$. \\
  Our~description~is~exhaustive~for~genera~1~and~2.
  As previously observed \cite{Anoevo,DPP}, evolution at $T\neq -1$ is not fully smooth:
  it switches abruptly at the boundaries between different regions.
  We reveal that this happens also at the boundary between thin and thick knots,
  moreover, the thick-knot domain is further stratified.
  For thin knots the two eigenvalues $1$ and $\lambda = q^2 T$, governing the evolution,
  are the standard $T$-deformation of the eigenvalues of the $R$-matrix 1 and $-q^2$.
  However, in thick knots' regions extra eigenvalues emerge, and they are powers
  of the ``naive'' $\lambda$, namely, they are equal to $\lambda^2, \dots, \lambda^g$.
  From point of view of frequencies, i.e. logarithms of eigenvalues, this is frequency doubling
  (more precisely, frequency multiplication) -- a phenomenon typical for non-linear dynamics.
  Hence, our observation can signal a hidden non-linearity of superpolynomial evolution.
  To give this newly observed evolution a short name, note that when $\lambda$ is pure phase
  the contributions of $\lambda^2, \dots, \lambda^g$ oscillate ``faster'' than the one of $\lambda$.
  Hence, we call this type of evolution ``nimble''.
}

\begin{document}
\today
\maketitle

{\section{Introduction} \label{sec:introduction}
  { 
    It is well-known that HOMFLY-PT polynomials \cite{knotpols}
    possess \textit{evolution structure} \cite{IMMMfe}-\cite{KNTZ}.
    This has a simple explanation within the modernized Reshetikhin-Turaev (MRT) formalism
    \cite{RT},
    and the evolution eigenvalues are actually those of the ${\mathcal{R}}$-matrix in
    the relevant representations.
    There is no known {\it a priori} reason to expect such structure in superpolynomials,
    defined in a very different way \cite{Kho,DGR} (see, however, \cite{DMMSS} and \cite{Do3}).
    Still,  in attempts to find a refined version of MRT,
    one  can try to {\it observe} a similar structure for Khovanov polynomials empirically
    -- and is immediately gratified:
    evolution was already proved to persist for the series of torus and twist knots
    \cite{Do12,Anoevo,DPP}.
    For example,   the $n$-dependence of {\it reduced} Khovanov invariant is of the form
    \be \label{eq:evolution-ansatz}
    \mathcal{X}^{\text{Torus}[2,n]} = C_1 \lambda_1^n + C_2 \lambda_2^n
    \ee
    and for positive odd $n$ it is actually
    \be
    \mathcal{X}^{\text{Torus}[2,n]}   \sim
    \frac{1 -q^2T + q^4T^2}{1-q^2T}
    - \frac{(q^2T)^{n+1}}{1-q^2T} = 1 + q^4T^2\cdot\frac{1-(q^2T)^{n-1}}{1-q^2T}
    \ee
    i.e. an explicitly {\it positive polynomial}.
    Switching to negative $n$ makes this expression explicitly {\it negative},
    and positivity is restored by insertion of additional overall factor $(-T)$.
    Additional simple modifications are needed for even $n$ and for unreduced invariants,
    which might look like a minor issue and, indeed, in this particular example
    can be explained away by a simple requirement that invariants
    remain {\it positive} and {\it minimal} for all $n$.
    However, as one considers more and
    more general knot/link families it becomes increasingly clear that there
    is more to the story.

    In this paper we look at a rather representative family
    of pretzel knots (see Section~\ref{sec:pretzel-knots-def} for a definition),
    which includes the entire twist and double-twist series,
    but only 2-strand sub-family of torus knots.
    Their evolution at HOMFLY-PT and, partly, superpolynomial
    levels was described in detail in \cite{IMMMfe}-\cite{KNTZ} and \cite{GMMMS-pretzel,pretzel-rest}.
    Here we study the evolution of Khovanov polynomials for this family.
    We immediately see that parameter space has rich, even puzzling, chamber structure:
    transitions between the chambers
    (an analog of changing the sign or parity of evolution parameter in 2-strand torus case)
    \textit{cannot} be fully explained by the positivity requirement
    (this line of thought, however, does not break completely, see
    Remarks~\ref{rem:positivity-comment-1}~and~\ref{rem:positivity-comment-2}).
    Before going into details we briefly outline what happens.
  }

  {\subsection{The problem} \label{sec:intro-the-problem}

    { 
      In the region where all winding parameters are positive, {\it reduced}
      Khovanov polynomials for pretzel knots
      (not  links! -- see Section~\ref{sec:reduced-khovanovs-links}) of genus $g$ are given
      by the general formula
      \begin{align} \label{eq:p*-reduced-knots-khovanov} 
        \boxed {
          \mathcal{X}_{n_i > 0}^{\text{knots}}
          =
          \frac{q }{s^g \qtTwo}
          \frac{1}{[2]_{qt}^{g+1}}
          \brackets{ \prod_{i=0}^g \Big( 1 +  {[3]_{qt}}  \brackets{q^2 T}^{n_i}\Big)
            + [3]_{qt} \prod_{i=0}^g \Big( 1 -   \brackets{q^2 T}^{n_i}\Big)}
        }
      \end{align}
      Here $s = \sqrt{-T}$ and $qt$-numbers are
      $[n]_{qt} = \frac{(sq)^{n} - (sq)^{-n}}{sq - (sq)^{-1}}\sim \frac{1-(-q^2T)^n}{1+q^2T}$
      (note that they are themselves {\it not} positive,
      but combine in an intricate way inside \eqref{eq:p*-reduced-knots-khovanov} to give a positive result
      -- see Remarks~\ref{rem:positivity-comment-1}~and~\ref{rem:positivity-comment-2}).
      This formula, however, is too simple: modulo trivial normalization coefficient
      it can be obtained just by the {\it change of variables}
      $q^2 \rightarrow (-T) \cdot q^2,\ A^2 \rightarrow (-T) \cdot q^4$
      from  the arborescent formula \cite{GMMMS-pretzel,pretzel-rest,arbor}
      for the corresponding HOMFLY-PT  polynomial --
      reflecting the fact
      that all knots in this region are homologically \textit{thin} \cite{katlas}.
      That is, the arborescent formula \cite{arbor} survives in this case
      not only the generalization to superpolynomial,
      but also the reduction to Khovanov ($N=2$) polynomials,
      which are defined and calculated in an absolutely different way.
    }

    { 
      However, as one goes out of the positive octant, one immediately encounters discrepancies.
      The simplest example is provided by the pair of 3-strand torus knots,
      $\text{Torus}[3,4]$ and $\text{Torus}[3,5]$, which are still pretzels
      (there are no more torus pretzels except these two and the 2-strand series).
      Indeed, of  the five terms in the reduced Khovanov polynomial
      \begin{align} \label{eq:torus-3-4-reduced-khovanov}
        \mathcal{X}\brackets{\text{Torus}[3,4]} \equiv
        \mathcal{X}\brackets{\text{Pretzel}[3,3,-2]} = & \ q^{13} T^6+q^9 T^4+q^7 T^3+q^7 T^2+q^3 T
      \end{align}
      only three are reproduced by the formula \eqref{eq:p*-reduced-knots-khovanov},
      provided one  multiplies it by an extra $(-T)$:
      \begin{align} \label{eq:red-kh-discrepancy}
        (\ref{eq:p*-reduced-knots-khovanov}) \ \Longrightarrow \    q^{13} T^6+q^9 T^4+q^3 T
      \end{align}
      And, as a rule, the discrepancy gets worse and worse as one moves away from the positive octant
      -- the presented example is by no means unique.
    }

    { 
      Taken in isolation, this is not so big a problem and not even a surprise.
      Indeed, Poincare polynomials of differential complexes,
      of which Khovanov polynomial is an example, usually behave much worse than corresponding Euler characteristics.
      But if one remembers the \textit{context}, which exists on $T = -1$ level, then discrepancy
      \eqref{eq:red-kh-discrepancy} is very important. Indeed, at $T = -1$, the analog of
      \eqref{eq:p*-reduced-knots-khovanov} has deep representation theory connections;
      it is made of so-called Racah matrix \cite{Racah}. This immediately allows one to generalize
      $\eqref{eq:p*-reduced-knots-khovanov}_{T = -1}$ to the \textit{colored} case, simultaneously
      revealing its connection to Chern-Simons \cite{CS} theory.

      If one ever hopes to have similarly rich context at $T \neq -1$ level, then
      understanding, or at least {\it taming}, this naive \textit{breakdown}
      of \eqref{eq:p*-reduced-knots-khovanov} is crucial, and this is precisely what we do in the present paper.

      Another point of interest is that proper description of the $T \neq -1$ structure
      may shed some light on the use of the topological string formalism
      to calculate refined knot polynomials. So far, this was understood only in the example
      of double Hopf link \cite{MMhopf}.

    }
  }

  { 
    \subsection{The main results} \label{sec:intro-main-results}
    In this paper we look at Khovanov polynomials for low genus pretzel knots\footnote{
      We do concrete calculations with the help of wonderful programs by Dror Bar-Natan and his collaborators
      \cite{BarNatan,BarNatan-fast,BarNatanprog}, with our own set of wrappers \cite{cl-vknots}.
      We also changed $q \rightarrow 1/q$, $T \rightarrow 1/T$ and
      chose a very specific framing
      (see Section~\ref{sec:unorientability-and-framing})
      in which the symmetry between different winding parameters $n_i$
      in \eqref{eq:p*-reduced-knots-khovanov} is manifest.
    }
    and find the following loosely related structures:

    {\subsubsection{Nimble evolution in exceptional regions}
      The abovementioned $\text{Pretzel}[3,3,-2]$ is near the tip of a special region in the parameter space
      \begin{align} \label{eq:exceptional-region-def}
        n_g \leq 0, \ \ n_i > -n_g, \ i = 0 \ ..\  g-1
      \end{align}
      where reduced Khovanov polynomials receive \textbf{unsymmetric} correction term
      \begin{align} \label{eq:red-kh-intro-unsym-corr}
        \boxed {
          \frac{q}{s^g} \frac{(1 + T)}{\qtTwo^{g}} \prod_{j = 0}^{g-1} \brackets{1 + \qtThree (q^2 T)^{n_j + n_g}}
      } \end{align}
      in which $n_g$ is distinguished and plays a special role.
      This term, of course, vanishes at $T = -1$.
      There are a few regions, shaped similarly
      to \eqref{eq:exceptional-region-def}, with more-or-less analogous
      kind of correction terms.

      The most prominent feature of \eqref{eq:red-kh-intro-unsym-corr} is that the dependence on $n_g$
      is very different from dependence on other windings $n_i$. $(q^2 T)^{n_g}$ occurs in \textit{each and every bracket}.
      Cumulative effect of these extra eigenvalues in all the brackets is that in the preferred direction
      evolution occurs \textit{faster} than would be naively expected. We call this phenomenon ``nimble evolution''.

      For arbitrary genus this is definitely not the whole story,
      but in Section~\ref{sec:ultra-low-genera} we present
      the details of what we understand so far.

      For genera $g=1$ and $g=2$, however, this description of reduced Khovanov polynomials
      for knots is {\bf exhaustive} and complete --
      the only deviation from (\ref{eq:p*-reduced-knots-khovanov}) are
      correction terms \eqref{eq:red-kh-arb-g-except-charged} and \eqref{eq:g-2-neutral-evo},
      analogous to \eqref{eq:red-kh-intro-unsym-corr},
      appearing in ``exceptional'' regions, shaped by inequalities \eqref{eq:exceptional-charged-shape}
      and \eqref{eq:exceptional-neutral-shape}.
      Only in these exceptional regions does one encounter {\it thick} knots,
      i.e. such knots (as opposed to {\it thin} knots) whose Khovanov polynomial contains $(q,T)$-monomials
      that do not lie on the cricical diagonals of the Newton plane
      (see Section~1.1 in \cite{bib:Manion-3-strand-pretzels} and referenced therein).
      While for thin knots Khovanov polynomial can be obtained from the respective Jones polynomial
      by simple change of variables, for thick knots one cannot do it,
      and this is what makes thin-thick knot distinction so important.
    }

    {\subsubsection{Unreduced polynomials can be restored from reduced ones}
      \label{sec:intro-unreduced-reduced-interplay}
      For genus 2 the {\bf unreduced} Khovanov polynomials
      can be recovered from reduced ones by adding simple corrections
      (see Section~\ref{sec:unreduced-khovanovs-knots}).
      They also change abruptly between strata,
      but inside each stratum they depend only
      on the planar diagram's \textit{unorientability}
      (see Sections~\ref{sec:unreduced-khovanovs-knots} and \ref{sec:unorientability-and-framing}).
    }

    {\subsubsection{Link polynomials have similar structure}
      Unreduced Khovanov polynomials for {\bf links}
      are not very much different from unreduced Khovanov polynomials
      for knots: they have simple extra correction terms
      that depend on the mutual linking numbers of the components and unorientability
      (see Section~\ref{sec:unorientability-and-framing}) of the planar diagram.
      Still, the structure of these terms is so different from arborescent structure
      \eqref{eq:p*-reduced-knots-khovanov}
      that joining links with different number of connected components
      into one evolution series (as was done in \cite{DPP})
      is more confusing than illuminating
      (see Section~\ref{sec:reduced-khovanovs-links}).

      We completely leave the question of structures present
      in reduced Khovanov polynomials for links
      out of this paper. This is mainly because
      reduced Khovanov polynomials for links require a different point of view:
      to any given link one associates not just one, but the whole bunch of polynomials,
      one for each choice of marked connected component.
    }

    \bigskip

    \noindent In this paper we present \textit{an interpretation} of the extensive
    experimental data on Khovanov polynomials.
    Of course, what we really want in the future, is to do \textit{prediction}:
    to write down formulas similar to \eqref{eq:red-kh-intro-unsym-corr} beforehand
    from some kind of guiding principle
    and then check that they indeed give Khovanov polynomials, calculated with help
    of their explicit definition.

    We conclude by discussing the meaning and limitations of our results
    and pointing further directions
    in Section~\ref{sec:conclusions}.
  }
}

{\section{Pretzel knots} \label{sec:pretzel-knots-def}
  \newcommand\theBraid[1]{
    { 
      \linethickness{0.5mm}
      \put(0,0){\line(0,1){60}}
      \put(0,0){\line(1,0){15}}
      \put(0,60){\line(1,0){15}}
      \put(15,0){\line(0,1){60}}
      \put(3,26){$#1$}
    }
  }

  Recall that pretzel knot of genus $g$ is a certain kind of knot that
  can be drawn on a genus $g$ surface.
  It consists of $g+1$ 2-strand braids, with winding numbers $n_0$ through $n_g$,
  respectively, which are joined, as shown on the picture.
  
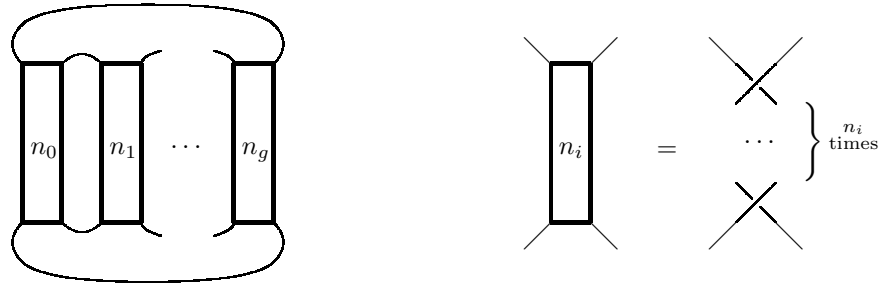
\begin{align} \label{eq:pretzel-planar-diagram}
    \begin{picture}(300,150)(-20,-50)
      \put(0,0){ 
        \thicklines
        \put(0,0){\theBraid{n_0}}
        \put(15,0){ 
          \put(0,60){\thinlines \qbezier(0, 0)(7.5, 7.5)(15, 0)}
          \put(0,0){\thinlines \qbezier(0, 0)(7.5, -7.5)(15, 0)}
        }
        \put(30,0){\theBraid{n_1}}
        \put(45,0){ 
          \put(0,60){\thinlines \qbezier(0, 0)(0,3)(7.5, 5)}
          \put(0,0){\thinlines \qbezier(0, 0)(0,-3)(7.5, -5)}
        }
        \put(56,28){$\dots$}
        \put(80,0){ 
          \put(0,60){\thinlines \qbezier(0, 0)(0,3)(-7.5, 5)}
          \put(0,0){\thinlines \qbezier(0, 0)(0,-3)(-7.5, -5)}
        }
        \put(80,0){\theBraid{n_g}}
        \put(0,60){ 
          \thinlines
          \qbezier(0,0)(-7.5,7.5)(0,15)
          \qbezier(95,0)(102.5,7.5)(95,15)
          \qbezier(0,15)(8,22.5)(47.5,22.5)
          \put(95,0){\qbezier(0,15)(-8,22.5)(-47.5,22.5)}
        }
        \put(0,0){ 
          \thinlines
          \qbezier(0,0)(-7.5,-7.5)(0,-15)
          \qbezier(95,0)(102.5,-7.5)(95,-15)
          \qbezier(0,-15)(8,-22.5)(47.5,-22.5)
          \put(95,0){\qbezier(0,-15)(-8,-22.5)(-47.5,-22.5)}
        }
      }
      \put(200,0){ 
        \put(0,0){
          \theBraid{n_i}
          \thinlines
          \put(0,0){\line(-1,-1){10}}
          \put(15,0){\line(1,-1){10}}
          \put(0,60){\line(-1,1){10}}
          \put(15,60){\line(1,1){10}}
        }
      }
      \put(240,25){$=$}
      \put(270,0){ 
        \thinlines
        \put(0,0){\line(-1,-1){10}}
        \put(15,0){\line(1,-1){10}}
        \put(0,60){\line(-1,1){10}}
        \put(15,60){\line(1,1){10}}
        \put(0,0){\prePositiveCrossing}
        \put(0,45){\prePositiveCrossing}
        \put(3,30){$\dots$}
        \put(25,28){$\Bigg\}$}
        \put(35,30){$\substack{n_i \\ \text{times}}$}
      }
    \end{picture}
  \end{align}
  In order to define framing (see Section~\ref{sec:unorientability-and-framing}),
  it is important to choose a particular planar projection, and for pretzel knots
  we always have in mind this one.

  Depending on parities of windings $n_i$,
  pretzel planar diagram \eqref{eq:pretzel-planar-diagram} can be either a knot or a link.
  A diagram is a knot, when either:
  \begin{itemize}
  \item one of the windings is even, and all the rest are odd
  \item genus $g$ is even and all the windings are odd
  \end{itemize}
  In the former case the ``even'' braid has to be antiparallel, while all ``odd'' braids
  are parallel. In the latter case all the braids are antiparallel
  
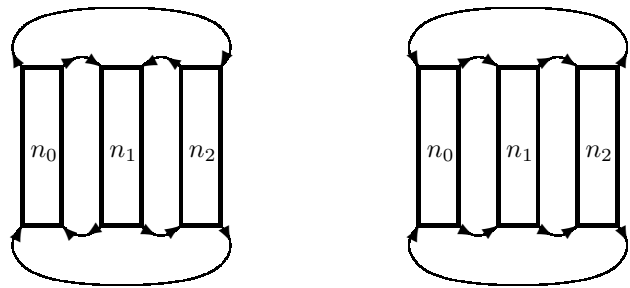
\begin{align}
    \begin{picture}(300,150)(-20,-50)
      \put(0,0){ 
        \thicklines
        \put(0,0){\theBraid{n_0}}
        \put(15,0){ 
          \put(0,60){\thinlines \qbezier(0, 0)(7.5, 7.5)(15, 0)}
          \put(0,0){\thinlines \qbezier(0, 0)(7.5, -7.5)(15, 0)}
        }
        \put(0,0){ 
          \put(0,0){\vector(1,2){0}}
          \put(15,0){\put(0,0){\vector(-1,1){0}}}
          \put(0,60){
            \put(0,0){\put(-4,6){\vector(-1,2){0}}}
            \put(15,0){\put(6,4){\vector(2,1){0}}}
          }
        }
        \put(30,0){\theBraid{n_1}}
        \put(45,0){ 
          \put(0,60){\thinlines \qbezier(0, 0)(7.5, 7.5)(15, 0)}
          \put(0,0){\thinlines \qbezier(0, 0)(7.5, -7.5)(15, 0)}
        }
        \put(30,0){ 
          \put(0,0){\put(-6,-4){\vector(-3,-2){0}}}
          \put(15,0){\put(6,-4){\vector(2,-1){0}}}
          \put(0,60){
            \put(0,0){\vector(3,-2){0}}
            \put(15,0){\put(0,0){\vector(-2,-1){0}}}
          }
        }
        \put(60,0){\theBraid{n_2}}
        \put(60,0){ 
          \put(0,0){\vector(3,2){0}}
          \put(15,0){\put(4,-6){\vector(1,-2){0}}}
          \put(0,60){
            \put(0,0){\put(-6,4){\vector(-3,2){0}}}
            \put(15,0){\put(0,0){\vector(-1,-2){0}}}
          }
        }
        \put(0,60){ 
          \thinlines
          \qbezier(0,0)(-7.5,7.5)(0,15)
          \qbezier(75,0)(82.5,7.5)(75,15)
          \qbezier(0,15)(8,22.5)(37.5,22.5)
          \put(75,0){\qbezier(0,15)(-8,22.5)(-37.5,22.5)}
        }
        \put(0,0){ 
          \thinlines
          \qbezier(0,0)(-7.5,-7.5)(0,-15)
          \qbezier(75,0)(82.5,-7.5)(75,-15)
          \qbezier(0,-15)(8,-22.5)(37.5,-22.5)
          \put(75,0){\qbezier(0,-15)(-8,-22.5)(-37.5,-22.5)}
        }
      }
      \put(150,0){ 
        \thicklines
        \put(0,0){\theBraid{n_0}}
        \put(15,0){ 
          \put(0,60){\thinlines \qbezier(0, 0)(7.5, 7.5)(15, 0)}
          \put(0,0){\thinlines \qbezier(0, 0)(7.5, -7.5)(15, 0)}
        }
        \put(0,0){ 
          \put(0,0){\vector(1,2){0}}
          \put(15,0){\put(6,-4){\vector(2,-1){0}}}
          \put(0,60){
            \put(0,0){\vector(1,-2){0}}
            \put(15,0){\put(6,4){\vector(2,1){0}}}
          }
        }
        \put(30,0){\theBraid{n_1}}
        \put(45,0){ 
          \put(0,60){\thinlines \qbezier(0, 0)(7.5, 7.5)(15, 0)}
          \put(0,0){\thinlines \qbezier(0, 0)(7.5, -7.5)(15, 0)}
        }
        \put(30,0){ 
          \put(0,0){\vector(3,2){0}}
          \put(15,0){\put(6,-4){\vector(2,-1){0}}}
          \put(0,60){
            \put(0,0){\vector(3,-2){0}}
            \put(15,0){\put(6,4){\vector(2,1){0}}}
          }
        }
        \put(60,0){\theBraid{n_2}}
        \put(60,0){ 
          \put(0,0){\vector(3,2){0}}
          \put(15,0){\put(4,-6){\vector(1,-2){0}}}
          \put(0,60){
            \put(0,0){\vector(3,-2){0}}
            \put(15,0){\put(4,6){\vector(1,2){0}}}
          }
        }
        \put(0,60){ 
          \thinlines
          \qbezier(0,0)(-7.5,7.5)(0,15)
          \qbezier(75,0)(82.5,7.5)(75,15)
          \qbezier(0,15)(8,22.5)(37.5,22.5)
          \put(75,0){\qbezier(0,15)(-8,22.5)(-37.5,22.5)}
        }
        \put(0,0){ 
          \thinlines
          \qbezier(0,0)(-7.5,-7.5)(0,-15)
          \qbezier(75,0)(82.5,-7.5)(75,-15)
          \qbezier(0,-15)(8,-22.5)(37.5,-22.5)
          \put(75,0){\qbezier(0,-15)(-8,-22.5)(-37.5,-22.5)}
        }
      }
    \end{picture}
  \end{align}

  For the purposes of this paper we will call the former pretzel knots (that have exactly one
  antiparallel braid) \textit{charged} and the latter pretzel knots \textit{neutral},
  since the former ones have non-zero unorientability (see Section~\ref{sec:unorientability-and-framing}),
  while the latter ones do not.

  It is crucial to distinguish charged and neutral pretzel knots, since,
  as we shall see in Section~\ref{sec:ultra-low-genera},
  starting from genus $g = 2$ in some regions evolution formulas
  for these two types of pretzel knots do differ.
}

{\section{Reduced Khovanov polynomials} \label{sec:ultra-low-genera}
  { 
    In this section we present the evolution formulas
    for reduced Khovanov polynomials. We go incrementally, from the simpler
    formulas valid in some regions of the parameter space, to more and more
    complicated formulas. }

  { 
    Here, unless otherwise specified, index $i$ runs from $0$ to $g$,
    index $J$ is some distinguised index
    (and in this case the region considered is the union of regions for all
    possible choices of $J$).
    Here, and in the following sections as well, $\lambda$ is equal to $q^2 T$:
    \begin{align}
      \lambda := q^2 T
    \end{align}
  }

  { 
    The simplest possible formula is
    \begin{align} \label{eq:red-kh-arb-g-bulk-g}
      \mathcal{X}_{\text{bulk}_g}^{\text{knots}}
      = &
      \frac{q (-T)}{(q^{-1} - q T)}
      \frac{1}{(q^{-1} - q T)^{g+1}}
      \brackets{ \prod_{i=0}^g \Big( 1 +  {[3]_{qt}}  \brackets{q^2 T}^{n_i}\Big)
        + [3]_{qt} \prod_{i=0}^g \Big( 1 -   \brackets{q^2 T}^{n_i}\Big)
      },
    \end{align}
    which is valid in the region
    \begin{align} \label{eq:bulk-g-shape}
      \text{bulk}_g \ & : \ (n_i > 0) \opor (n_J = 0 \opand n_{i \neq J} > 0) \opor (n_J = -1 \opand n_{i \neq J} > 1)
    \end{align}
    The motivation behind the region's name will become clear in a second.
    The formula \eqref{eq:red-kh-arb-g-bulk-g} is straightforwardly obtained from the HOMFLY polynomial
    with help of change of variables $q^2 \rightarrow (-T) \cdot q^2,\ A^2 \rightarrow (-T) \cdot q^4$.
    This is to be expected, since all the knots in this region are \textit{alternating} and, hence, homologically thin
    (which precisely means they can be restored from respective HOMFLY with the substitution).

    The formula \eqref{eq:red-kh-arb-g-bulk-g} for sure cannot be true on the entire windings space,
    since, as one tries to apply it outside the $\text{bulk}_g$ region, it stops giving positive answer
    (see Remarks~\ref{rem:positivity-comment-1}~and~\ref{rem:positivity-comment-2}).
  }

  { 
    The failure of positivity of \eqref{eq:red-kh-arb-g-bulk-g} is, in fact, cured
    in a very easy way in a number of regions, which we denote $\text{bulk}_a$, $a = -g, -g + 2, \dots, g - 2, g$.
    The shape of these regions is, in general, complicated
    (at least so far we were unable to find a generic description of their shape by some inequalities),
    but one of the regions -- $\text{bulk}_{-g}$ -- is the antipode of $\text{bulk}_{g}$:
    \begin{align} \label{eq:bulk-min-g-shape}
      \text{bulk}_{-g} \ & : \ (n_i < 0) \opor (n_J = 0 \opand n_{i \neq J} < 0) \opor (n_J = 1 \opand n_{i \neq J} < -1)
    \end{align}

    \noindent The correct formula in bulk-regions is
    \begin{align} \label{eq:red-kh-arb-g-bulk-a}
      \mathcal{X}_{\text{bulk}_a}^{\text{knots}} = (-T)^{\frac{a - g}{2}} \mathcal{X}_{\text{bulk}_g}^{\text{knots}}
    \end{align}

    \begin{remark}
      In $g = 1$ case, there are just two regions $\text{bulk}_1$ and $\text{bulk}_{-1}$, which are larger
      than in general case, namely
      \begin{align}
        \text{bulk}_1 \ & : \ n_0 + n_1 > 0
        \\ \notag
        \text{bulk}_{-1} \ & : \ n_0 + n_1 < 0
      \end{align}
      i.e. they span the whole parameter space (the diagonal $n_0 = n_1$ contains only links).
      Note that there is no separate restriction on $n_0$ and $n_1$ -- just on their sum,
      because for $g=1$ it is easy to rewrite \eqref{eq:red-kh-arb-g-bulk-g} to depend manifestly only on $n_0 + n_1$.
    \end{remark}
    
\begin{remark}
Note that the mirror symmetry, which is a fundamental property of the Khovanov polynomials, presents here in the form
\begin{align}
\mathcal{X}(\text{Pretzel}[n_0,\ldots,n_g])(q,T)=q^2\mathcal{X}(\text{Pretzel}[-n_0,\ldots,-n_g])(q^{-1},T^{-1}),
\label{eq:red-mirr}
\end{align}
where an extra factor of $q^2$ is a due to the peculiarity of the definition of the \textit{reduced} polynomials~\cite{BarNatan}. One can explicitly verify that (\ref{eq:red-mirr}) indeed relates the p and m versions of all our evolution formulas.
\end{remark}


\begin{remark} \label{rem:positivity-comment-1}
One can observe that all our evolution formulas are in fact assembled from the elementary factors of the three kinds,
\begin{align}
\arraycolsep=0mm
\begin{array}{lcl}
f_n(\lambda)&=&\lambda^{-n}\frac{1-\lambda^n}{1-\lambda}=\frac{\lambda^{-n}-1}{1-\lambda},\\[3mm] 
g_n(\lambda)&=&(\lambda^{-1}-1+\lambda)\lambda^{n}f_n(\lambda)=\lambda^{-1}-\lambda^{n-1}+\lambda\Big(\frac{1-\lambda^{n-2}}{1-\lambda}+\lambda^{n-2}+\lambda^{n-1}\Big)=
\lambda^{-1}+\lambda^{n-1}f_{n-2}(\lambda)+\lambda^n,\\[2mm] F_n(\lambda)&=&\lambda^{-n}\frac{1+[3]_{qT}\lambda^n}{1-\lambda}=\frac{\lambda^{-n}-\lambda^{-1}+1-\lambda}{1-\lambda}=\lambda+\frac{\lambda^{-n+1}-1}{1-\lambda}
. 
\end{array}
\label{facs}
\end{align}
These factors are (Laurent) polynomials in $\lambda$ for any integer $n$. Moreover, $F_n(\lambda)$ and $f_n(\lambda)$ are positive (negative) polynomials for $n>0$ ($n<0$), and $g_n(\lambda)$ is a positive (negative) polynomials for $n>1$ ($n<-1$).  All these polynomials are \textit{almost} proportional to ordinary quantum numbers $[n]_q$ with $\lambda$ on the place of $q$ (see the explicit examples in app.~\ref{app:pols}). In addition, (\ref{facs}) satisfy certain relations (see app.~~\ref{app:facs}) that allow one to rewrite the evolution formulas as \textit{explicitly} positive polynomials.

In particular, one can rewrite $g=1$ answer (\ref{eq:red-kh-arb-g-bulk-g}) in the form 
\begin{align} 
\mathcal{X}^{n_0,n_1}=q^4T\cfrac{\lambda^{n_0+n_1}}{\lambda-1}\Big(F_{n_0}F_{n_1}-\lambda^{-n_1}f_{n_0}g_{n_1}\Big)
=q^2\lambda^{n_0+n_1}F_{n_0+n_1}(\lambda),\label{eq:red-kh-g-1-pdec}
\end{align}
so that it depends only on $n_0+n_1$ (as it should) and literally coincides with the standard Khovanov polynomial (under the considerations from the beginning of sec.~\ref{sec:intro-main-results}) of the knot $\text{Torus}[2,n_1+n_0]\sim\text{Pretzel}(n_0,n_1)$~\cite{Anoevo}.

Similarly, algebraic manipulations with (\ref{eq:red-kh-arb-g-bulk-g}) allow one to rewrite it as   
   \begin{align} 
       \mathcal{X}_{\text{bulk}_2}^{\text{knots}}
       = q^5T\cfrac{\lambda^{n_0+n_1}}{1-\lambda}\Big(F_{n_0}\!(\lambda)F_{n_1}\!(\lambda)F_{n_2}\!(\lambda)-f_{n_0}\!(\lambda)f_{n_1}\!(\lambda)\lambda^{-n_2}g_{n_2}\!(\lambda)\Big)=\nn\\
      =q^3\lambda^{n_0+n_1}\Big(F_{n_0}\!(\lambda)F_{n_1}\!(\lambda)+F_{n_0}\!(\lambda)g_{n_2}\!(\lambda)+f_{n_1}\!(\lambda)g_{n_2}\!(\lambda)\Big).
       \label{eq:red-kh-g-2-pdec}
     \end{align}
The last expression is an \textit{explicitly} positive polynomial for $n_0>1$, $n_1>0$, and $n_2>1$. Moreover, one can find several equivalent forms of (\ref{eq:red-kh-arb-g-bulk-g}) with their own domains of explicit positivity (or negativity), so that the union of these domains is exactly the union of all the $\text{bulk}_a$ regions.

Analogues of (\ref{eq:red-kh-g-2-pdec}) for other (not bulk-region) evolution formulas for $g=2$ are presented below. The higher genera evolution formulas reveal very similar structures, but we postpone this for the upcoming work on systematic analysis of these cases.
\end{remark}


    \noindent For $g > 1$ bulk-regions do not span the whole space, but they still do take
    a significant (say, greater than $1/2$) fraction of its volume.

    \begin{remark}
      In the bulk-regions it doesn't matter, whether knot is charged or neutral
      -- formula \eqref{eq:red-kh-arb-g-bulk-a} interpolates between both possibilities.
    \end{remark}

  }

  { 
    Formula \eqref{eq:red-kh-arb-g-bulk-a} also does not provide correct answers on the whole parameter space.
    Already for $g = 2$ one has torus knots $\text{Torus}[3,4]$ and $\text{Torus}[3,5]$
    for which there is a discrepancy (typeset in bold)
    \begin{align} \label{eq:torus-3-4-and-3-5-discrepancies}
      \mathcal{X}\brackets{\text{Torus}[3,4]} \equiv
      \mathcal{X}\brackets{\text{Pretzel}[3,3,-2]} = & \ q^{13} T^6+q^9 T^4+q^3 T+\boldsymbol{q^7 T^3 + q^7 T^2}
      \\ \notag
      \mathcal{X}\brackets{\text{Torus}[3,5]} \equiv
      \mathcal{X}\brackets{\text{Pretzel}[5,3,-2]} = & \
      q^{17} T^8 + q^{13} T^6 - q^5 T^2 + \boldsymbol{q^{11} T^5 + q^{11} T^4  + q^7 T^3 + q^7 T^2 + q^5 T^2 + q^5 T}
    \end{align}
    We see that in case of $\text{Torus}[3,5]$ the mismatch is more severe:
    the naive bulk answer does not give sign-definite polynomial at all!
    

    Nevertheless, extra bold terms in both $\text{Torus}[3,4]$ and $\text{Torus}[3,5]$ are successfully accounted for by the following corrected formulas
    \be
    \arraycolsep=0.5mm
    \begin{array}{ccccrcl}
    \mathcal{X}_{\mathrm{bulk_\mathbf{-2}}}^{\text{Pretzel}[3,3,-2]}
          &=&\boldsymbol{-T}\mathcal{X}_{\mathrm{bulk_\mathbf{2}}}^{\text{Pretzel}[3,3,-2]}
          &=&-q^7\boldsymbol{T^3}&+&T\!q^3\big(1\!+\!q^4T^2\big)\big(1\!+\!q^6T^3\big), 
    \\
              &&\mathcal{X}^{\text{Pretzel}[3,3,-2]}&=&q^7\boldsymbol{T^2}&+&Tq^3\big(1\!+\!q^4T^2\big)\big(1\!+\!q^6T^3\big), 
    \end{array}
    \label{ex332}\ee
    and
    \be
    \arraycolsep=0mm
    \begin{array}{ccccrcl}
\mathcal{X}_{\mathrm{bulk_{\mathbf{-2}}}}^{\text{Pretzel}([5,3,-2])}
&=&\boldsymbol{-T}\mathcal{X}_{\mathrm{bulk_{\mathbf{2}}}}^{\text{Pretzel}([5,3,-2])} 
&=&
    -q^5\boldsymbol{T^2}\big(1\!+\!q^2T\!+\!q^{6}T^3\big)&+&q^7T^3\big(1\!+\!q^4T^2\big)\big(1\!+\!q^6T^3\big), 
    \\[2mm]
    &&\mathcal{X}^{\text{Pretzel}([5,3,-2])}&=&q^5\boldsymbol{T}\big(1\!+\!q^2T\!+\!q^{6}T^3\big)&+&q^7T^3\big(1\!+\!q^4T^2\big)\big(1\!+\!q^6T^3\big).
    \label{ex532}\end{array}
    \ee
    The bold type indicates above the $T$ powers that are different in the bulk and actual evolution formulas. These factors are  responsible for the bold terms in (\ref{eq:torus-3-4-and-3-5-discrepancies}) and for the cancellation of the negative term for $\text{Torus}[3,5]$.are successfully accounted for by the following corrected formulas

    Generally, the evolution formulas 
    \begin{align} \label{eq:red-kh-arb-g-except-charged}
             \mathcal{X}_{\text{pExceptCharged}}^{\text{knots}} & =
             (-T) \mathcal{X}_{\text{bulk}_g}^{\text{knots}} +
             q (1 + T) \frac{1}{(q^{-1} - q T)^g} \prod_{i \neq J} \brackets{1 + \qtThree \lambda^{n_i + n_J}}
             \\ \notag
             \mathcal{X}_{\text{mExceptCharged}}^{\text{knots}} & =
             (-T)^{-1} \mathcal{X}_{\text{bulk}_{-g}}^{\text{knots}} +
             q (1 + T^{-1}) \frac{(-T)^g}{(q^{-1} - q T)^g} \prod_{i \neq J} \brackets{1 + \qtThree \lambda^{n_i + n_J}}.
           \end{align}
    are valid, respectively, in positive exceptional charged region
    and negative exceptional charged region, whose shape is
    \begin{align} \label{eq:exceptional-charged-shape}
      \braces{\substack{\text{positive} \\ \text{exceptional} \\ \text{charged}}}
       \ & : \ n_J \text{ is even} \opand n_J \leq 0 \opand n_{i \neq J} > - n_J; \ \ \vec{n} \notin \text{bulk}_g
      \\ \notag
      \braces{\substack{\text{negative} \\ \text{exceptional} \\ \text{charged}}}
      \ & : \ n_J \text{ is even} \opand n_J \geq 0 \opand n_{i \neq J} < - n_J; \ \ \vec{n} \notin \text{bulk}_{-g}
    \end{align}
    That is, each of the exceptional charged regions consists of $g+1$ subregions,
    corresponding to the choice of special direction $J = 0 \dots g+1$.
    Moreover, all the knots in the region are \textit{charged}, since $n_J$ is even,
    justifying the name of these regions.

    \begin{remark}
      Crucial feature of the evolution formulas \eqref{eq:red-kh-arb-g-except-charged}
      (and of the formulas~\eqref{eq:g-2-neutral-evo} below)
      is that eigenvalue $\lambda$ corresponding to the chosen
      preferred direction $J$ enters all the brackets of the correction term, while eigenvalues
      corresponding to other, non-preferred, directions each enter precisely one bracket.
      Hence, if we consider evolution w.r.t just $n_J$, with other $n_i$ fixed, then it occurs
      \textit{faster} (resulting in extra terms in \eqref{eq:torus-3-4-and-3-5-discrepancies})
      than would be naively guessed.
      We call this \textbf{nimble evolution} and hope to study in the future how it manifests itself in the regions
      of the parameter space we haven't covered so far.
    \end{remark}


  \begin{remark}
  The positive polynomial decomposition over elementary factors (\ref{facs}) in the case of $g=2$, e.g., for the first of formulas (\ref{eq:red-kh-arb-g-except-charged}) is
  \begin{align} \label{eq:red-kh-g-2-ex-ch-pdec}
  \mathcal{X}_{\text{pExceptCharged}}^{\text{knots}}=q^3 \lambda^{n_0+n_1+n_2}\big(\lambda^{n_2}F_{n_0\!+\!n_2}F_{n_1\!+\!n_2}+\boldsymbol{T}F_{n_2}g_{n_2}\big).
         \end{align}
  Remarkably, bulk formula (\ref{eq:red-kh-arb-g-bulk-g})$\equiv$(\ref{eq:red-kh-g-2-pdec}) is recovered from (\ref{eq:red-kh-g-2-ex-ch-pdec}) if one substitutes the bold $\boldsymbol{T}$ with $-1$, just as we have seen in explicit examples (\ref{ex332}, \ref{ex532}).
  \end{remark}

  
  }

  { 
    
    The odd, or neutral, counterpart of the exceptional regions
    \begin{align} \label{eq:exceptional-neutral-shape}
      \braces{\substack{\text{positive} \\ \text{exceptional} \\ \text{neutral}}}
      \ & : \ {n_J \text{ is odd} \opand n_J \leq 1 \opand n_{i \neq J} > - n_J}; \ \ \vec{n} \notin \text{bulk}_g
      \\ \notag
      \braces{\substack{\text{negative} \\ \text{exceptional} \\ \text{neutral}}}
      \ & : \ {n_J \text{ is odd} \opand n_J \geq -1 \opand n_{i \neq J} < - n_J}; \ \ \vec{n} \notin \text{bulk}_{-g}
    \end{align}
    requires for a more complicated description, which we present here only for $g=2$,
    \begin{align} \label{eq:g-2-neutral-evo}
             \mathcal{X}_{\text{pExceptNeutral}}^{\text{knots}} & =
             (-T) \mathcal{X}_{\text{bulk}_2}^{\text{knots}} +
             q (1 + T) \frac{\qtThree}{(q^{-1} - q T)^2}\brackets{1 + \lambda^{n_2+1}}\brackets{1 + \lambda^{n_2-1}}
             \\ \notag
             \mathcal{X}_{\text{mExceptNeutral}}^{\text{knots}} & =
             (-T)^{-1} \mathcal{X}_{\text{bulk}_{-2}}^{\text{knots}} +
             q (1 + T)\frac{\qtThree (-T)^2}{(q^{-1} - q T)^2}\brackets{1 + \lambda^{n_2+1}}\brackets{1 + \lambda^{n_2-1}}.
           \end{align}
    On very shallow level, the structure of \eqref{eq:g-2-neutral-evo} is still similar to
    \eqref{eq:red-kh-arb-g-except-charged}. That is, there is still one preferred direction $J$,
    and evolution in this direction is nimble. And the correction terms still vanish at $T = -1$.
    But understanding the structure of \eqref{eq:g-2-neutral-evo}
    on a deeper level, as well as the systematic analysis of higher genera, is the subject for future research.
    In particular, for $g > 2$ ``bulk'' and ``exceptional'' regions from above do not span
        the whole parameter space -- there are additional regions, where the dependence
        of the Khovanov polynomial is still to be described.

        \begin{remark} \label{rem:positivity-comment-2}
        Decomposition over positive polynomial (\ref{facs}), e.g., in the first case, is
        \begin{align} \label{eq:red-kh-g-2-ex-n-pdec}
        \mathcal{X}_{\text{pExceptNeutral}}^{\text{knots}}=q^3\lambda^{n_0+n_1+n_2}\Big(\lambda^{n_2}( F_{n_0\!+\!n_2}F_{n_1\!+\!n_2-1}\!+\lambda^{-1}\!F_{n_0\!+\!n_2}\!)+\boldsymbol{T}f_{n_2-1}g_{n_2+1}\Big).
               \end{align}       
        Again, the substitution of $-1$ for the bold $\boldsymbol{T}$ turns (\ref{eq:red-kh-g-2-ex-n-pdec}) into bulk formula (\ref{eq:red-kh-arb-g-bulk-g})$\equiv$(\ref{eq:red-kh-g-2-pdec}). 
        \end{remark}
        
  }

  { 
    \begin{remark}
      For $g=2$ the regions $\text{bulk}_{\pm 2}$, $\text{bulk}_{0}$
      and positive and negative exceptional charged and neutral regions span the entire space
      (the $\text{bulk}_{0}$ is the \textit{complement} of all other regions).
      Hence, for $g=2$ formulas \eqref{eq:red-kh-arb-g-bulk-a}, \eqref{eq:red-kh-arb-g-except-charged}
      and \eqref{eq:g-2-neutral-evo} provide \textbf{complete description} for reduced Khovanov polynomials' evolution.
    \end{remark}

    \begin{remark}
      The double-braid knots,
      instrumental in finding a relation between inclusive and exclusive Racah matrices
      \cite{MMMS-racah-and-hidden-integrability,arbor},
      are embedded into $\text{bulk}_2$ region for $g=2$ as $\text{Pretzel}[n_0, -1, n_2]$.
      This is a weak hint that evolution formula \eqref{eq:red-kh-arb-g-bulk-g} should
      be at the core of the (hypothetical) homological analog of the arborescent calculus.
    \end{remark}

    \begin{remark}
      While charged exceptional regions, indeed, contain only charged knots,
      the neutral exceptional regions contain \textit{both} charged and neutral knots.
      Namely, they contain those charged knots for which the preferred direction $J$
      does not coincide with the direction, which has even winding.
      For instance, a charged pretzel knot $\text{Pretzel}[5, -3, 4]$ belongs
      to positive exceptional neutral region with $J = 1$ (the distinguished direction),
      while its only antiparallel braid corresponds to winding $n_2 = 4$.
    \end{remark}

    \begin{remark}
      One may wonder whether choosing the preferred direction in exceptional regions
      is consistent with \textbf{topological invariance}.
      Note that topological invariance implies only invariance of the answers
      w.r.t cyclic permutation of the winding numbers, for example
      \begin{align}
        \text{Pretzel}[5, -3, 4] = \text{Pretzel}[4, 5, -3] = \text{Pretzel}[-3, 4, 5]
      \end{align}
      That is, to reproduce these answers one needs to use formula \eqref{eq:g-2-neutral-evo} with
      \textit{different} $J = 1$, $2$ and $0$, respectively.
    \end{remark}
  }
}

{\section{Relation between reduced and unreduced Khovanov polynomials} \label{sec:unreduced-khovanovs-knots}
  It turns out that in each stratum of the parameter space unreduced polynomials
  can be recovered from the reduced ones.
  For genus 2 the description below is exhaustive,
  while for higher genera we don't yet know what happens in some of the regions.

  The relation betwen reduced ($\mathcal{X}$) and unreduced ($X$)
  polynomials is particularly simple in bulk-regions
  \begin{align}
    X_{\text{bulk}_a}^{\text{knots}}
    = \frac{(1 + q^4 T)}{q^2 (1 + q^2 T)} \mathcal{X}_{\text{bulk}_a}^{\text{knots}}
    + \frac{q^a (1 + T)}{(1 + q^2 T)} \lambda^{\text{unorientability}},
  \end{align}
  where $\text{unorientability}$ is a simple combinatorial quantity associated to a planar diagram
  and is defined in Section~\ref{sec:unorientability-and-framing}.

  In exceptional charged regions it is slightly more complicated, for instance,
  \begin{align}
    X_{\text{pExceptCharged}}^{\text{knots}}
    = & \frac{(1 + q^4 T)}{q^2 (1 + q^2 T)} \mathcal{X}_{\text{pExceptCharged}}^{\text{knots}}
    \\ \notag
    + & \frac{q^{g-1} (1 + T)}{(1 + q^2 T)} \lambda^{\text{unorientability}}
    + \frac{q^{1-g} (1 + T)(1+q^4 T)}{(1 + q^2 T) T} \lambda^{2 n_J} \lambda^{\text{unorientability}}
  \end{align}
  Though each individual correction term is very simple,
  their generic structure is not clear at the moment:
  more research is needed to clarify the issue.

  Since we, in any case, don't have a generic description, this section is very sketchy,
  but from what we observe so far, the jumps in unreduced and reduced Khovanov homology
  occur \textit{together} -- chambers for reduced and unreduced polynomials are the same.
}

{\section{Unorientability and framing} \label{sec:unorientability-and-framing}
  Unorientability is defined as follows. Consider checkerboard coloring of the planar diagram
  (where we've denoted colored regions with black circles):
  
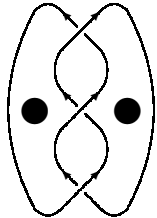
\begin{align}
    \begin{picture}(300,80)(-130,0)
      \put(0,0){
        \put(0,0){\crossingPosUp}
        \put(0,15){
          \qbezier(0,0)(-5,7.5)(0,15)
          \put(15,0){\qbezier(0,0)(5,7.5)(0,15)}
        }
        \put(0,30){\crossingPosUp}
        \put(0,45){
          \qbezier(0,0)(-5,7.5)(0,15)
          \put(15,0){\qbezier(0,0)(5,7.5)(0,15)}
        }
        \put(0,60){\crossingPosUp}
        \put(0,0){
          \qbezier(0,0)(-5,-5)(-10,0)
          \put(-10,0){\qbezier(0,0)(-20,37.5)(0,75)}
          \put(0,75){\qbezier(0,0)(-5,5)(-10,0)}
        }
        \put(15,0){
          \qbezier(0,0)(5,-5)(10,0)
          \put(10,0){\qbezier(0,0)(20,37.5)(0,75)}
          \put(0,75){\qbezier(0,0)(5,5)(10,0)}
        }
        \put(-10,37){\circle*{10}}
        \put(25,37){\circle*{10}}
      }
    \end{picture}
    \nn
  \end{align}
  Out of the two possible choices we choose the one that doesn't contain an infinite region.
  Now, contributions of different types of crossings to the unorientability are
  \begin{align}
    \begin{picture}(300,25)(-130,0)
      \put(-130,0){
        \crossingPosUp
        \put(0,7.5){\circle*{5}}
        \put(15,7.5){\circle*{5}}
        \put(23,5){$=\ 0$}
      }
      \put(-70,0){
        \crossingNegUp
        \put(0,7.5){\circle*{5}}
        \put(15,7.5){\circle*{5}}
        \put(23,5){$=\ 0$}
      }
      \put(-10,0){
        \crossingPosUp
        \put(7.5,0){\circle*{5}}
        \put(7.5,15){\circle*{5}}
        \put(23,5){$=\ +1$}
      }
      \put(50,0){
        \crossingNegUp
        \put(7.5,0){\circle*{5}}
        \put(7.5,15){\circle*{5}}
        \put(23,5){$=\ -1$}
      }
    \end{picture}
  \end{align}

  Throughout the paper, we use a very particular choice of framing (with respect to Bar-Natan's conventions).
  This is needed in order to restore the symmetry between different windings
  $n_i$, even though some of them correspond to parallel braids and others
  correspond to antiparallel braids.
  Namely, the required framing factor is simply $(T q^3)$ to the power of unorientability,
  which for pretzel knots is equal to sum of windings of parallel-oriented braids:
  \begin{align}
    \braces{\substack{\text{framing} \\ \text{factor}}}
    = \brackets{T q^3}^{\text{unorientability}} = \prod_{i : \substack{\text{parallel} \\ \text{braid}}} \brackets{T q^3}^{n_i}
  \end{align}
}

{\section{Unreduced Khovanov polynomials for pretzel links} \label{sec:reduced-khovanovs-links}
  If we consider links, not just knots, and try to interpolate between different answers
  for unreduced Khovanov polynomials then for the $\text{bulk}_g$ region we would get
  \begin{align} \label{eq:khovanov-p*-unreduced}
    X_{\text{bulk}_g}
    = & \frac{(1 + q^4 T) (-T)^{g/2}}{(1 + q^2 T)(1 - q^2 T)} \frac{1}{[2]_{qt}^{g+1}}
    \brackets{\prod_{i=0}^g \brackets{1 + [3]_{qt} \lambda^{n_i}}
      + [3]_{qt} \prod_{i=0}^g \brackets{1 - \lambda^{n_i}}
    }
    \\ \notag
    + & \frac{(T + 1)}{2 (1 + q^2 T)} \brackets{
      \prod_{i = 0}^g \brackets{\frac{q}{2} + \frac{q}{2}(-1)^{n_i} + q (q^2 T)^{n_i}}
      + \prod_{i = 0}^g \brackets{\frac{q}{2} + \frac{q}{2}(-1)^{n_i} - q (q^2 T)^{n_i}}
    }
    \\ \notag
    + & \frac{(T + 1)}{(1 + q^2 T)} \prod_{i = 0}^g \brackets{-\frac{q}{2} + \frac{q}{2}(-1)^{n_i}}
  \end{align}

  It is clear that the answer  changes abruptly when one changes the number of link components
  (i.e. the number of windings $n_i$ that are even).

  Namely, if we have an $M$-component link, then the correction w.r.t the naive arborescent answer
  is
  \begin{align} \label{eq:unreduced-khovanov-link-corrections}
    \Delta X_{\text{bulk}_g} = &
    \ \frac{(T + 1)}{2 (1 + q^2 T)}
    q^{g+1} \lambda^{\text{unorientability}} \brackets{
      \prod_{C_i < C_j} (1 + \lambda^{2 \ \text{lk}(C_i, C_j)})
      + (-1)^{g+1 - M} \prod_{C_i < C_j} (1 - \lambda^{2 \ \text{lk}(C_i, C_j)})
    }
    \\ \notag
    + & \ \frac{(T + 1)}{(1 + q^2 T)} (-q)^{g + 1} \delta_{M,1} \delta_{\text{unorientability},0},
  \end{align}
  where we've written it in the form that has a chance to generalize \textit{beyond} the pretzel knots.
  Here unorientability of a planar diagram is as in Section~\ref{sec:unorientability-and-framing},
  $\text{lk}(C_i, C_j)$ is the linking number of the link components $C_i$ and $C_j$,
  and products $\prod_{C_i < C_j}$ run over distinct pairs of link components.

  Overall, we see that corrections \eqref{eq:unreduced-khovanov-link-corrections} look very
  differently from the arborescent piece. Hence, rather than trying to find a formula
  that interpolates between knots and links (with varying number of components),
  it is much more fruitful to direct attention to formulas for links with \textit{fixed}
  number of components.
  The main focus of the present paper was on knots, but, hopefully, this section shows
  that answers for links with other number of components are only a little bit more complicated.

}

{\section{Different approaches to similar problems} \label{sec:related-work}
  Here we briefly review different papers, that are in some way related
  to what we do in this paper.

  {\subsection{Khovanov polynomials for genus 2 Prezel knots}
    An orthogonal research direction to our experimental approach
    consists in honest symbolic computation of Khovanov polynomials ``by hands'',
    i.e. in honestly \textit{deriving} formulas like
    \eqref{eq:p*-reduced-knots-khovanov} and \eqref{eq:red-kh-intro-unsym-corr},
    rather than getting them via interpolation.

    The key point here is that the Khovanov's complex for an open two strand braid has a simple and explicit description.
    Moreover, the complexes for the two strand braids can be multiplied
    (via the operation of so-called horizontal composition)
    so that a pretzel knot (or link) is obtained, and its Khovanov polynomial can be thus explicitly computed.
    This plan was gradually implemented for all genus two pretzel knots.
    Here are the relevant milestones.

    Pioneering takes on the problem relied in an essential way
    on the exact skein sequence and the differential expansion
    (which substitute the skein relations and the quantum group structure, respectilely).
    For quasi-alternating links, which constitute a large fraction of all links at genus two,
    this resulted in the general Theorem~4.5 of~\cite{bib:Lee:endomorphism-of-khovanov-invariant}
    for the unreduced polynomials.

    The next step was the explicit computation of unreduced Khovanov polynomials for several
    infinite series of non-quasi-alternating genus 2 pretzel links
    \cite{bib:Suzuki:khovanov-homology-and-rasmussen-invariant,
      bib:Starkson:khovanov-homology-of-ppq-pretzel-knots,
      bib:Qazaqzeh:khovanov-homology-three-column}.
    All these polynomials proved to be homologically thin,
    and thus similar to the polynomials of the alternating links.

    The remaining genus two pretzel links were captured in~
    \cite{bib:Manion:rational-Khovanov-homology-3-strand-pretzel-links}.
    The paper contains the general answer for the unreduced polynomial of a pretzel link.
    In particular, this answer explicitly shows that some families of the genus 2 pretzel links are homolgically thick,
    i.e., the corresponding Khovanov polynomials are not fully defined by other invariants.

    Hence, this cooperated research provides the complete list of the explicit formulas for the unreduced Khovanov polynomials for genus 2 pretzel links.
    Yet, the evolution formulas were never presented in a condensed and consice form in these papers,
    as we do in the present paper.
    This, we hope, is one of our main contributions to this development,
    and hopefully will give a clue on how to extend explicit description to higher genera.
  }

  { \subsection{Evolution formulas for Khovanov(-Rozansky) polynomials}
    { 
      The focused study of the evolution of Khovanov-Rozansky polynomials
      at finite $N$,
      to the best of our knowledge, was started in \cite{Anoevo}.
      There, the authors concentrated their attention on the case of torus knots,
      which, on one hand, allowed them to study Khovanov-Rozansky polynomials,
      and not just Khovanov ($N = 2$) ones, but on the other hand,
      concealed the full generality of the chamber structure
      -- there the chamber structure took the form of the breaking of the
      mirror symmetry.
      A very interesting aspect of the paper \cite{Anoevo} is that
      the main role is played not by the KR-polynomials themselves, but rather
      by finite difference equations, that these polynomials satisfy.
      In the present paper we do not comment on this approach at all,
      but this dual point of view is a potential source of many new insights.
    }

    { \subsection{Evolution formulas for double-braid knots}
      { 
        Fourth of all, the present paper is the development of \cite{DPP}.
        There, also, evolution for Khovanov polynomials (i.e. $N=2$) was studied for a concrete
        family of knots -- the double-braid knots (which authors called ``figure-eight-like'').
        The richness of the chamber structure for Khovanov polynomials was already observed there,
        moreover, answers were proven, not just guessed from computer experiments, as in the present paper.
        Pretzel knots, considered in the present paper, contain double braid ones,
        for example, as $\text{Pretzel}[a,b,1]$. An interesting feature of \cite{DPP} is that
        evolution formulas are written for knots and links jointly, which results
        in appearance of extra eigenvalue.
        Now, our analysis in Section~\ref{sec:reduced-khovanovs-links} suggests
        that this point of view is more confusing that it is fruitful -- it is much
        more instructive to consider links with different number of components as different evolution series.
      }
    }
  }

  {\subsection{Superpolynomials of torus knots\label{sec:superpolynomials of torus knots}}
    { 
      Other but closely related objects are superpolynomials for torus knots, studied in~\cite{DMMSS}.

      Superpolynomials are, roughly speaking, ``stable component'' of the Khovanov-Rozansky
      polynomials. Namely, if one studies Khovanov-Rozansky polynomials for any given knot
      for different ranks $N$ of the group, for $N > N^*$ (where $N^*$ depends on the knot)
      the dependence on $N$ becomes analytic -- polynomial stabilizes.
      In particular, at the level of superpolynomials evolution method works \textit{perfectly},
      what was further confirmed in the case of twist knots in \cite{evo}.
      Chambers with abrupt changes between them appeared in these considerations,
      but these changes  could be easily ignored in \cite{evo} by saying that
      evolution smoothly connects pure positive polynomials with pure negative ones --
      what is true in the  twist and torus cases.
      For the first time the {\it seriousness} of the chamber problem for superpolynomials
      was realized in the study of satellite knots in \cite{satel}.
      As we explain in the present paper, the problem is indeed very general, just in the case
      of pretzels it fully manifests itself for {\it finite} $N$.
      Thus chamber dependence can be considered as a kind of pronounced
      non-perturbative  phenomenon, which is strengthened beyond the large-$N$ (loop) expansion --
      and this is what we study in the present paper.
    }
  }

  { 
    \bigskip
    \noindent There are, of course, many more papers that are related to the present work
    in one way or another.
    We do not pretend to make a comprehensive review here
    -- we only mention results, which directly affected the motivations
    and content of the present paper.
  }
}

{\section{Conclusion and further directions} \label{sec:conclusions}
  In this paper we analyzed
  the explicit expressions for Khovanov polynomials
  for pretzel knots of low genera,
  obtained from computer experiments with the help of \cite{BarNatanprog}
  (with our custom set of wrappers, which make our life more convenient,
    but are not necessarily easy to read \cite{cl-vknots}),
  and, partly, from direct computations of \cite{bib:Manion-3-strand-pretzels}.

  We were mainly interested in the fate of the evolution formulas.
  We observed that chamber structure is very rich for this family of knots.
  While for some knots (alternating and quasi-alternating) evolution is very simple and
  just follows from evolution for HOMFLY-PT polynomials,
  for other knots (the {\bf thick} pretzel knots) there are non-trivial corrections.
  But, perhaps, the main surprise and good news
  is that our suggested formulas \eqref{eq:red-kh-arb-g-except-charged} and \eqref{eq:g-2-neutral-evo}
  are still of the shape that is \textit{comparable} to naive answer \eqref{eq:red-kh-arb-g-bulk-g}.
  This gives a hope that some homological generalization of MRT-formalism,
  or even arborescent calculus, is, indeed, possible.
  Before, the only \textit{multiparametric} family of knots, for which such generalization
  was constructed (on the level of superpolynomials \cite{DMMSS}) were torus knots,
  i.e. generalized was the celebrated Rosso-Jones formula \cite{RJ}.

  Apart from generalizing our formulas to higher genera, another obvious research route
  would be to understand their quadruply-graded homology analogues \cite{GGS,NawOb}.

  Finally, the study of (q,t)-deformed pretzel formulas may be helpful
  in developing explicit formulas for the Racah matrices (quantum 6j-symbols) themselves.
  So far even at $T = -1$ their description is far from being complete
  (see \cite{hypergeom} for current state of art)
  and it well may be that some aspects become clearer as one goes to $T \neq -1$.

  So far the picture we present is complete only for genera 1 and 2, while already for genus 3
  there are regions, where  the form of the evolution is still obscure, hence,
  we can not insist that
  corrections are always as tame as \eqref{eq:red-kh-arb-g-except-charged} or \eqref{eq:g-2-neutral-evo}.
  Something more wild is still not excluded.
  Our work is continuing in these directions.
}

\section*{Acknowledgements}

This work was funded by the Russian Science Foundation (Grant No.16-12-10344)

\appendix

\section{Elementary constituents of Pretzel Khovanov polynomials\label{app:facs}}

Here we discuss elementary building blocks~(\ref{facs}) of the Pretzel Khovanov polynomials in little more details. We repeat the definition for the sake of convenience,
\be
&f_n(z)=\cfrac{z^{-n}-1}{1-z},\hspace{2.5cm} g_n(z)=(z^{-1}-1+z)z^{n}f_n(z),\label{appfacs}\\ &F_n=z+\cfrac{z^{-n+1}-1}{1-z}=1-z^{-1}+f_n(z)=z^{-n}\big(1-z^{-1}+g_n(z)\big).\\[4mm]
\nn
\ee
These factors satisfy there are the sum formulas that extend similar formulas for the quantum numbers look like
\be
f_{n_1+n_2}(z)=f_{n_1}(z)+z^{-n_1}f_{n_2}(z)=z^{-n_2}f_{n_1}(z)+f_{n_2}(z),\\
g_{n_1+n_2}(z)=z^{n_2}g_{n_1}(z)+g_{n_2}(z)=
g_{n_1}(z)+z^{n_1}g_{n_2}(z),\\
F_{n_1+n_2}(z)=F_{n_1}(z)+z^{-n_1}f_{n_2}(z),\ \ z^{n_2}F_{n_1+n_2}(z)=F_{n_1}(z)+g_{n_2}(z),
\label{sum}\ee
Relations between different factors (\ref{appfacs}), together with sum formulas (\ref{sum}), allow one to derive positive polynomials decompositions (\ref{eq:red-kh-g-1-pdec},\ref{eq:red-kh-g-2-pdec},\ref{eq:red-kh-g-2-ex-ch-pdec},\ref{eq:red-kh-g-2-ex-n-pdec}), as well as similar decompositions in other cases, including the higher genera evolution formulas. In particular, formulas for the Pretzel subfamilies in app.~\ref{app:pols} are obtained just in this way.

Formulas (\ref{facs},\ref{sum}) are valid for any integer $n$, $n_1$, $n_2$. Unlike them, the Laurent polynomials in $z$ obtained for particular values of $n$ look differently depending on the $n$ sign. Namely, 
\be
\begin{array}{lllll}
f_n(z)=\sum\limits_{i=1}^{n}z^{-i},& F_n=1+\sum\limits_{i=2}^{n}z^{-i}, & n>0;&
g_n(z)=z^{-1}+\sum\limits_{i=1}^{n-2}z^{i}+z^{n}, & n>1;\\[4mm]
f_n(z)=-\sum\limits_{i=0}^{-n-1}z^{i},& F_n=-z^{-1}-\sum\limits_{i=1}^{-n-1}z^{i}, &n<0;&
g_n(z)=-1-\sum\limits_{i=1}^{-n-1}z^{-i}-z^{n-1}, & n<-1.
\end{array}\label{facsdec}
\ee
I.e., (\ref{facs}) are fully positive or negative polynomials for the most positive or negative values of the integer $n$, respectively. 
One should treat separately the exceptional cases, when the factors are zero or sign indefinite,
\be
f_0(z)=g_0(z)=0,\ \ g_{-1}(z)=-1+z^{-1}-z^{-2},\ \ g_{1}(z)=z-1+z^{-1},\ \ F_0(z)=1-z^{-1}. 
\ee
In all cases, $f_n(z)$, $F_n(z)$ and $g_{-n}(z)$ contain only negative powers of $z$ if $n\ge 0$, and the $z^{-1}$ term followed by only positive powers of $z$ if $n\le 0$.  

\section{Explicit form of the unreduced Khovanov polynomials for the particular subfamilies of the genus 2 pretzel knots\label{app:pols}}
$\boxed{n_0=n_1=3,\ \ n_2=n}$
\nopagebreak

$
\arraycolsep=0.5mm
\begin{array}{|c|c|c|l|l|}
\hline\rule{0pt}{5mm}
n&\multicolumn{2}{c|}{q^{-3}\mathcal{X}^{3,3,n}(z,T)}&F_3\!\!=1\!+\!z^{\!\!-2}\!+\!z^{\!\!-3}&f_3\!\!=\!\!z^{\!\!-1}\!+\!z^{\!\!-2}\!+\!z^{\!\!-3}\\
\hline\hline\rule{0pt}{5mm}
\ldots&&
\ldots&\ \ \ldots&\ \ \ldots\\
\cline{1-1}\cline{3-5}\rule{0pt}{5mm}
6&&z^{12}F_3F_9\!+\!z^6f_3g_6&F_9\!\!=\!\!1\!+\!z^{\!\!-2}\!+\!z^{\!\!-3}+\!\!...\!\!+\!z^{\!\!-9}&
g_6\!\!=\!\!z^{\!\!-1}\!+\!z\!+\!z^2\!+\!z^3\!+z^4\!+\!z^6\\
\cline{1-1}\cline{3-5}\rule{0pt}{5mm}
5&z^6\big(\!F_3f_3\!+\!(\!F_3\!+\!f_3\!)g_n\!\big)\!\!=\!\!&z^{11}F_3F_8\!+\!z^6f_3g_5&
F_8\!\!=\!\!1\!+\!z^{\!\!-2}\!+\!z^{\!\!-3}+\!\!...\!\!+\!z^{\!\!-8}&
g_5\!\!=\!\!z^{\!\!-1}\!+\!z\!+\!z^2\!+\!z^3\!+\!z^5\\
\cline{1-1}\cline{3-5}\rule{0pt}{5mm}
4&\left(\!z^6\!+\!2z^4\!+\!2z^3\!+\!z^2\!+\!2z\!+\!1\!\right)&z^{10}F_3F_7\!+\!z^6f_3g_4&
F_7\!\!=\!\!1\!+\!z^{\!\!-2}\!+\!z^{\!\!-3}+\!\!...\!\!+\!z^{\!\!-7}&
g_4\!\!=\!\!z^{\!\!-1}\!+\!z\!+\!z^2\!+\!z^4\\
\cline{1-1}\cline{3-5}\rule{0pt}{5mm}
3&\!+\!\left(\!z^6\!+\!z^5\!+\!2z^4\!+\!2z^3\!\right)g_n&z^{9}F_3F_6\!+\!z^6f_3g_3&
F_6\!\!=\!\!1\!+\!z^{\!\!-2}\!+\!z^{\!\!-3}+\!\!...\!\!+\!z^{\!\!-6}&
g_3\!\!=\!\!z^{\!\!-1}\!+\!z\!+\!z^3\\
\cline{1-1}\cline{3-5}\rule{0pt}{5mm}
2&\!\!=\!\!z^{n\!+\!6}F_3F_{n\!+\!3}\!+\!z^{n\!+\!1}f_3g_n&z^{8}F_3F_5\!+\!z^6f_3g_2&
F_5\!\!=\!\!1\!+\!z^{\!\!-2}\!+\!z^{\!\!-3}+\!\!...\!\!+\!z^{\!\!-5}&
g_2\!\!=\!\!z^{\!\!-1}\!+\!z^2\\
\hline\rule{0pt}{5mm}
1&\multicolumn{2}{c|}{z^7F_3F_4+z^3g_3}&F_4\!\!=\!\!1\!+\!z^{\!\!-2}\!+\!z^{\!\!-3}\!+\!z^{\!\!-4}
&g_3\!\!=\!\!z^{\!\!-1}\!+\!z\!+\!z^3 
\\[2mm]
\hline\rule{0pt}{5mm}
0&\multicolumn{2}{c|}{z^6F_3^2}&F_3\!\!=1\!+\!z^{\!\!-2}\!+\!z^{\!\!-3}& 
\\[2mm]
\hline\rule{0pt}{5mm}
-\!1&\multicolumn{2}{c|}{z^3(\!F_2f_1+F_1\!)}
& 
F_2\!=\!1\!+\!z^{\!-\!2},f_1\!=\!z^{\!-\!1}\!\!,F_1\!\!=\!\!1& 
\\[2mm]
\hline\rule{0pt}{5mm}
-\!2&z^{2n+6}\!F_{3\!+\!n}^2\!+\!\boldsymbol{T}z^{n+6}\!F_{n}g_{n} &
z^{-3}F_{1}^2+\boldsymbol{T}zF_{\!-\!2}g_{\!-\!2} 
&
F_1\!=\!1,\ -\!F_{\!-2\!}\!=\!z^{\!-1}\!+\!z&-\!g_{\!-2}\!=\!1\!+\!z^{-3}
\\
\hline\rule{0pt}{5mm}
-\!3&\multicolumn{2}{c|}{ \boldsymbol{T}z^{-1}+z^5\!\big(\!f_2g_{\!-\!3}+g_{\!-\!2}\!\big)}
&f_2\!=\!z^{\!-\!1}\!\!+\!z^{\!-\!2}&\!-\!g_{\!-\!3}\!=\!1\!+\!z^{\!-\!2}\!+\!z^{\!-\!4} 
\\[2mm]
\hline\rule{0pt}{5mm}
-\!4&{\boldsymbol{\!-T}}\big(\!z^{n\!+\!6}F_3F_{n\!+\!3}\!+\!z^{n\!+\!1}f_3g_n\!\big)
&{\boldsymbol{\!-\!T}}\!\big(\!z^2F_3F_{\!\!-1}\!+\!z^6f_3g_{\!\!-4}\!\big)
&-\!F_{\!-1}\!=\!z^{\!-1}&\!-\!g_{\!-4}\!=\!1\!+\!z^{\!-2}\!+\!z^{\!-3}\!+\!z^{\!-5}\\ 
\cline{1-1}\cline{3-5}\rule{0pt}{5mm}
-\!5&
&{\boldsymbol{\!-\!T}}\!\big(\!z^3F_3F_{\!\!-2}\!+\!z^7f_3g_{\!\!-5}\!\big)&
-\!F_{\!-2}\!=\!z^{\!-1}\!+\!z&\!-\!g_{\!-5}\!=\!1\!+\!z^{\!-2}\!+\!\ldots\!+\!z^{\!-4}\!+\!z^{\!-6}
\\
\cline{1-1}\cline{3-5}\rule{0pt}{5mm}
-\!6&&{\boldsymbol{\!-\!T}}\!\big(\!z^4F_3F_{\!\!-3}\!+\!z^8f_3g_{\!\!-6}\!\big)&
-\!F_{\!-3}\!=\!z^{\!-1}\!+\!z\!+\!z^2&\!-\!g_{\!-6}\!=\!1\!+\!z^{\!-2}\!+\!\ldots\!+\!z^{\!-5}\!+\!z^{\!-7}
\\
\cline{1-1}\cline{3-5}\rule{0pt}{5mm}
\ldots&&
\ldots&\ \ \ldots&\ \ \ldots\\
\hline
\end{array}
$

$\boxed{n_0=n,\ n_2=3,\ \ n_2=-2}$
\nopagebreak

$
\arraycolsep=0.5mm
\begin{array}{|c|c|c|l|}
\hline\rule{0pt}{5mm}
n&
\multicolumn{2}{|c|}{q^{-3}\mathcal{X}^{n,3,-2}(z,T)}&
\!-\!F_{\!-\!2}\!=\!z^{\!\!-1}\!\!+\!z,\ \!-\!g_{\!-\!2}\!\!=1\!\!+\!z^{\!\!-3}\\
\hline\hline\rule{0pt}{5mm}
\ldots&
\ldots&\ \ \ldots&\ldots\\
\cline{1-1}\cline{3-4}\rule{0pt}{5mm}
7&z^{\!-5\!}F_{n\!-\!2}\!+\!\boldsymbol{T}z^{\!-\!1}F_{\!-\!2}g_{\!-\!2}
&z^{\!-3\!}F_5\!+\!\boldsymbol{T}z^{\!-\!1}F_{\!-\!2}g_{\!-\!2}
&F_{5}\!\!=\!\!1\!+\!z^{\!\!-2}\!+\!z^{\!\!-3}+\!\!z^{\!\!-4}\!\!+\!z^{\!\!-5}\\
\cline{1-1}\cline{3-4}\rule{0pt}{5mm}
5&&z^{\!-3\!}F_3\!+\!\boldsymbol{T}z^{\!-\!1}F_{\!-\!2}g_{\!-\!2}
&F_{3}\!\!=\!\!1\!+\!z^{\!\!-2}\!+\!z^{\!\!-3}\\
\cline{1-1}\cline{3-4}\rule{0pt}{5mm}
3&&z^{\!-3\!}F_1\!+\!\boldsymbol{T}z^{\!-\!1}F_{\!-\!2}g_{\!-\!2}
&F_{1}\!\!=\!1\\
\hline\rule{0pt}{5mm}
1&\boldsymbol{-\!T}z^{n+1}\big(\!F_{n}\!+\!z^2f_{2}g_{\!-\!3}\!\big)\!&
&f_3\!=\!z^{\!-\!1}\!\!+\!z^{\!-\!2}\!\!+\!z^{\!-\!3}\\
\hline\rule{0pt}{5mm}
-\!1&&\boldsymbol{T}^{\!-\!1}\big(\!z^2\bar{F}_1\!-\!\bar{f}_{2}\bar{g}_{-3}\!\big)
&\bar{F}_{1}\!\!=\!\!1
\\[2mm]
\cline{1-1}\cline{3-4}\rule{0pt}{5mm}
-\!3&\boldsymbol{T}^{\!-\!1}z^{\!-\!n}\big(\!z^3\bar{F}_{\!-\!n}\!-\!z\bar{f}_{3}\bar{g}_{-2}\!\big)
&\boldsymbol{T}^{\!-\!1}\big(\!z^5\bar{F}_3\!-\!z^3\bar{f}_{2}\bar{g}_{-3}\!\big)
&\bar{F}_{5}\!\!=\!\!1\!+\!z^{2}\!+\!z^{3}\\[2mm]
\cline{1-1}\cline{3-4}\rule{0pt}{5mm}
-\!5&&\boldsymbol{T}^{\!-\!1}\big(\!z^7\bar{F}_5\!-\!z^5\bar{f}_{2}\bar{g}_{-3}\!\big)
&\bar{F}_{5}\!\!=\!\!1\!+\!z^{2}\!+\!z^{3}+z^4+\!z^{5}\\
\cline{1-1}\cline{3-4}\rule{0pt}{5mm}
\ldots&&\ldots&\ldots\\
\hline\hline
\multicolumn{1}{|c}{\rule{0pt}{6mm}}&\multicolumn{2}{c|}{q^{-3}\mathcal{X}^{n,3,-2}(z,T)=q^{-2}q^{3}\mathcal{X}^{-n,2,-3}(z^{\!-\!1},T^{-\!1})}&
\bar{f}_2\!=\!z\!+\!z^{2},\ \ 
-\bar{g}_{\!-\!3}\!\!=1\!+\!z^{2}\!+\!z^4\\
\hline
\end{array}
$

\vspace{1cm}

$\boxed{n_0=n_1=5,\ \ n_2=n}$
\nopagebreak

$
\arraycolsep=0.5mm
\begin{array}{|c|c|c|l|l|}
\hline\rule{0pt}{5mm}
n&\multicolumn{2}{c|}{q^{-3}\mathcal{X}^{5,5,n}(z,T)}&F_5\!\!=1\!+\!z^{\!\!-2}\!+\!z^{\!\!-3}\!+\!z^{\!\!-4}
\!+\!z^{\!\!-5}
&f_5\!\!=\!\!z^{\!\!-1}\!+\!z^{\!\!-2}\!+\!z^{\!\!-3}\!+\!z^{\!\!-4}
\!+\!z^{\!\!-5}\\
\hline\hline\rule{0pt}{5mm}
\ldots&&
\ldots&\ \ \ldots&\ \ \ldots\\
\cline{1-1}\cline{3-5}\rule{0pt}{5mm}
2&z^{n+1}\big(\!z^9F_5F_{n\!+\!5}\!+\!f_5g_n\!\big)&z^{8}F_5F_{10}\!+\!z^6f_5g_2&
F_{10}\!\!=\!\!1\!+\!z^{\!\!-2}\!+\!z^{\!\!-3}+\!\!...\!\!+\!z^{\!\!-10}&
g_2\!\!=\!\!z^{\!\!-1}\!+\!z^2\\
\hline\rule{0pt}{5mm}
1&\multicolumn{2}{c|}{z^{11}F_{5}F_{6}+z^{5}g_{5}}
&F_6\!\!=\!\!1\!+\!z^{\!\!-2}\!+\!z^{\!\!-3}+\!\!...\!\!+\!z^{\!\!-6}&
g_5\!\!=\!\!z^{\!\!-1}\!+\!z\!+\!z^2\!+\!z^3\!+\!z^5 
\\[2mm]
\hline\rule{0pt}{5mm}
0&\multicolumn{2}{c|}{z^{10}F_5^2}&F_5\!\!=1\!+\!z^{\!\!-2}\!+\!z^{\!\!-3}\!+\!z^{\!\!-4}
\!+\!z^{\!\!-5}& 
\\[2mm]
\hline\rule{0pt}{5mm}
-\!1&\multicolumn{2}{c|}{z^{7}\big(F_{4}f_{3}+F_{3}\big)}
& \multicolumn{2}{l|}{
F_4\!\!=\!1\!+\!z^{\!\!-2}\!+\!z^{\!\!-3}\!+\!z^{\!\!-4}\!\!,
f_3\!\!=\!z^{\!\!-1}\!+\!z^{\!\!-2}\!+\!z^{\!\!-3}\!\!,
F_3\!\!\!=\!1\!+\!z^{\!\!-2}\!+\!z^{\!\!-3}\!} 
\\[2mm]
\hline\rule{0pt}{5mm}
-\!2&z^{2n+10}\!F_{n\!+\!5}^2\!+\!\boldsymbol{T}z^{n+10}\!F_{n}g_{n}&
z^6\!F_3^2\!+\!\boldsymbol{T}z^8\!F_{\!-\!2}g_{\!-\!2}&
-\!F_{\!-2\!}\!=\!z^{\!-1}\!+\!z,\ \ F_1\!=\!1&-\!g_{\!-2}\!=\!1\!+\!z^{-3}
\\
\hline\rule{0pt}{5mm}
-\!3&
\begin{array}{c}\rule{0pt}{5mm}
z^{2n+9}F_{n+5}(\!zF_{4+n}\!+\!1\!)\\
+\boldsymbol{T}z^{n+10}f_{n-1}g_{n+1}\end{array}
&z^4\!F_2F_1\!+\!z^3\!F_2 
\!+\!\boldsymbol{T}z^6\!f_{\!-\!4}g_{\!-\!2} 
&\multicolumn{2}{c|}{
F_2\!\!=\!1\!+\!z^{\!\!-2}\!\!,\ \
F_1\!\!=\!1,\ \ \ \ 
\!-\!f_{\!-\!4}\!\!=\!1\!+\!z\!+\!z^{2}\!+\!z^{3}\!}
\\[2mm]
\hline\rule{0pt}{5mm}
-\!4&z^{2n+10}\!F_{n\!+\!5}^2\!+\!\boldsymbol{T}z^{n+10}\!F_{n}g_{n}
&z^2F_1^2\!+\!\boldsymbol{T}z^6\!F_{\!-\!4}g_{\!-\!4}&
-\!F_{\!-4}\!=\!z^{\!-1}\!\!+\!z\!+\!z^2\!+\!z^3&\!-\!g_{\!-4}\!=\!1\!+\!z^{\!-2}\!+\!z^{\!-3}\!+\!z^{\!-5}\\ 
\hline\rule{0pt}{5mm}
-\!5&
\multicolumn{2}{|c|}{
\boldsymbol{T}z^{\!-1}+z^9\!\big(f_{4}g_{\!-\!5}+g_{\!-\!4}\big)}
&
f_4\!=\!z^{\!-\!1}\!\!+\!z^{\!-\!2}\!\!+\!z^{\!-\!3}\!\!+\!z^{\!-\!4}
&\!-\!g_{\!-5}\!=\!1\!+\!z^{\!-2}\!+\!\ldots\!+\!z^{\!-4}\!+\!z^{\!-6}
\\
\hline\rule{0pt}{5mm}
-\!6&{\boldsymbol{\!-\!T}}z^{n+1}\!\big(\!z^{9}F_5F_{\!n+5}\!+\!f_5g_n\!\big)
&{\boldsymbol{\!-\!T}}\!\big(\!z^4F_5F_{\!\!-1}\!+\!z^6f_5g_{\!\!-6}\!\big)&
-\!F_{\!-1}\!=\!-\!z^{\!-1}&\!-\!g_{\!-6}\!=\!1\!+\!z^{\!-2}\!+\!\ldots\!+\!z^{\!-5}\!+\!z^{\!-7}
\\
\cline{1-1}\cline{3-5}\rule{0pt}{5mm}
\ldots&&
\ldots&\ \ \ldots&\ \ \ldots\\
\hline
\end{array}
$

\end{document}